\documentclass[preprint, showpacs, longbibliography, prx,aps, superscriptaddress, amsmath, amssymb]{revtex4-1}
 \usepackage{graphicx}
 \usepackage{dcolumn}
 \usepackage{bm}
 \usepackage{color,ulem} 

\begin{document}
  \title{Coupling Bright and Dark Plasmonic Lattice Resonances}

  \author{S. R. K. Rodriguez}\email{s.rodriguez@amolf.nl}
  \affiliation{Center for Nanophotonics, FOM Institute AMOLF, c/o Philips Research Laboratories, High Tech Campus 4, 5656 AE Eindhoven, The Netherlands}

  \author{A. Abass}
   \affiliation{Department of Electronic and Information Systems (ELIS), Ghent University, Sint-Pietersnieuwstraat 41, B-9000 Ghent, Belgium}

  \author{B. Maes}
  \affiliation{Micro- and Nanophotonic Materials Group, Institut de Physique, University of Mons, Place du Parc 20, B-7000 Mons, Belgium}

  \author{O. T. A. Janssen}
  \address{Optics Research Group, Delft University of Technology, 2628 CJ Delft, The Netherlands}

  \author{G. Vecchi}
  \affiliation{Center for Nanophotonics, FOM Institute AMOLF, c/o Philips Research Laboratories, High Tech Campus 4, 5656 AE Eindhoven, The Netherlands}

  \author{J. G\'{o}mez Rivas}
  \affiliation{Center for Nanophotonics, FOM Institute AMOLF, c/o Philips Research Laboratories, High Tech Campus 4, 5656 AE Eindhoven, The Netherlands}
  \affiliation{Department of Applied Physics, Eindhoven University of Technology, P.O. Box 513, 5600 MB Eindhoven, The Netherlands}

  \date{\today}

\begin{abstract}
We demonstrate the coupling of bright and dark Surface Lattice
Resonances (SLRs), which are collective Fano resonances in 2D
plasmonic crystals. As a result of this coupling, a frequency
stop-gap in the dispersion relation of SLRs is observed. The
different field symmetries of the low and high frequency SLR bands
lead to pronounced differences in their coupling to free space
radiation. Standing waves of very narrow spectral width compared to
localized surface plasmon resonances are formed at the  high
frequency band edge, while subradiant damping onsets at the low
frequency band edge leading the resonance into darkness. We
introduce a coupled oscillator analog to the plasmonic crystal,
which serves to elucidate the physics of the coupled plasmonic
resonances and to estimate very high quality factors ($Q>700$) for
SLRs, which are the highest known for any 2D plasmonic crystal.

\end{abstract} 

\narrowtext \maketitle

Metallic nanoparticles supporting surface plasmon resonances allow
light to be localized in nanoscale volumes, thereby opening exciting
possibilities such as nanoscale control of
emitters~\cite{Muskens07}, large electromagnetic
enhancements~\cite{Aizpurua}, and nonlinear
nano-optics~\cite{Palomba&Novotny09}. Much attention has been given
to Localized Surface Plasmon Resonances (LSPRs), which arise in
individual particles when their conduction electrons are coherently
driven by an electromagnetic field. Although localized surface
plasmons may couple, their resonances are in general severely
broadened due to strong radiative damping and hence exhibit low
quality factors $Q$. A recent development in nanoplasmonics deals
with collective resonances in periodic arrays of metallic
nanostructures, or plasmonic crystals. Such arrays support Surface
Lattice Resonances (SLRs), which are collective resonances mediated
by diffractive coupling of localized plasmons. This coupling occurs
near the critical frequency when a radiating diffraction order
becomes evanescent, i.e., at the Rayleigh anomaly. SLRs were
introduced by Carron~\cite{Carron}, and the interest in this
phenomenon was revived by Schatz and co-workers with a series of
works on 1D and 2D arrays~\cite{Zou&Schatz04, Zou&Schatz05}.
However, the experimental observation of SLRs was elusive for many
years~\cite{Kall05}. Recent advances in nano-fabrication and in the
understanding of SLRs have allowed for their observation in periodic
arrays of nanostructures with different
geometries~\cite{Auguie&Barnes08, Crozier, Kravets08, Vecchi09,
Vecchi09b, Bitzer}. In contrast with LSPRs, SLRs possess much higher
$Q$s, and the associated polaritons can propagate over tens of unit
cells in the plasmonic crystal~\cite{Vecchi09b}. The relevance of
SLRs for enhanced, directional, and polarized light
emission~\cite{Vecchi09, Giannini10} and sensing~\cite{Offermans}
has been recently demonstrated. Although the coupling of surface
modes in periodic metallic structures has attracted much
interest~\cite{Barnes96b, Sauvan&Lalanne08, Ropers, Ghoshal&Kik09},
especially for its connection with frequency
stop-gaps~\cite{Barnes96}, coupled SLRs have not been discussed yet.

In this paper, we demonstrate the mutual coupling of SLRs and the
formation of a frequency stop-gap in the dispersion relation of these modes.
This coupling leads to a strong modification of the SLRs
characteristics, including the onset of subradiant damping in the
low frequency band, zero group velocity modes in the high frequency
band, and $Q$-factors for both bands which are amongst the highest
reported for any 2D plasmonic crystal. Our results set the basis for
controlling the dispersion of SLRs, and they open new possibilities
in sensing, enhanced spontaneous light emission, and lasing at the
band edges of SLR gaps.

\begin{figure}
\centerline{\scalebox{0.5}{\includegraphics{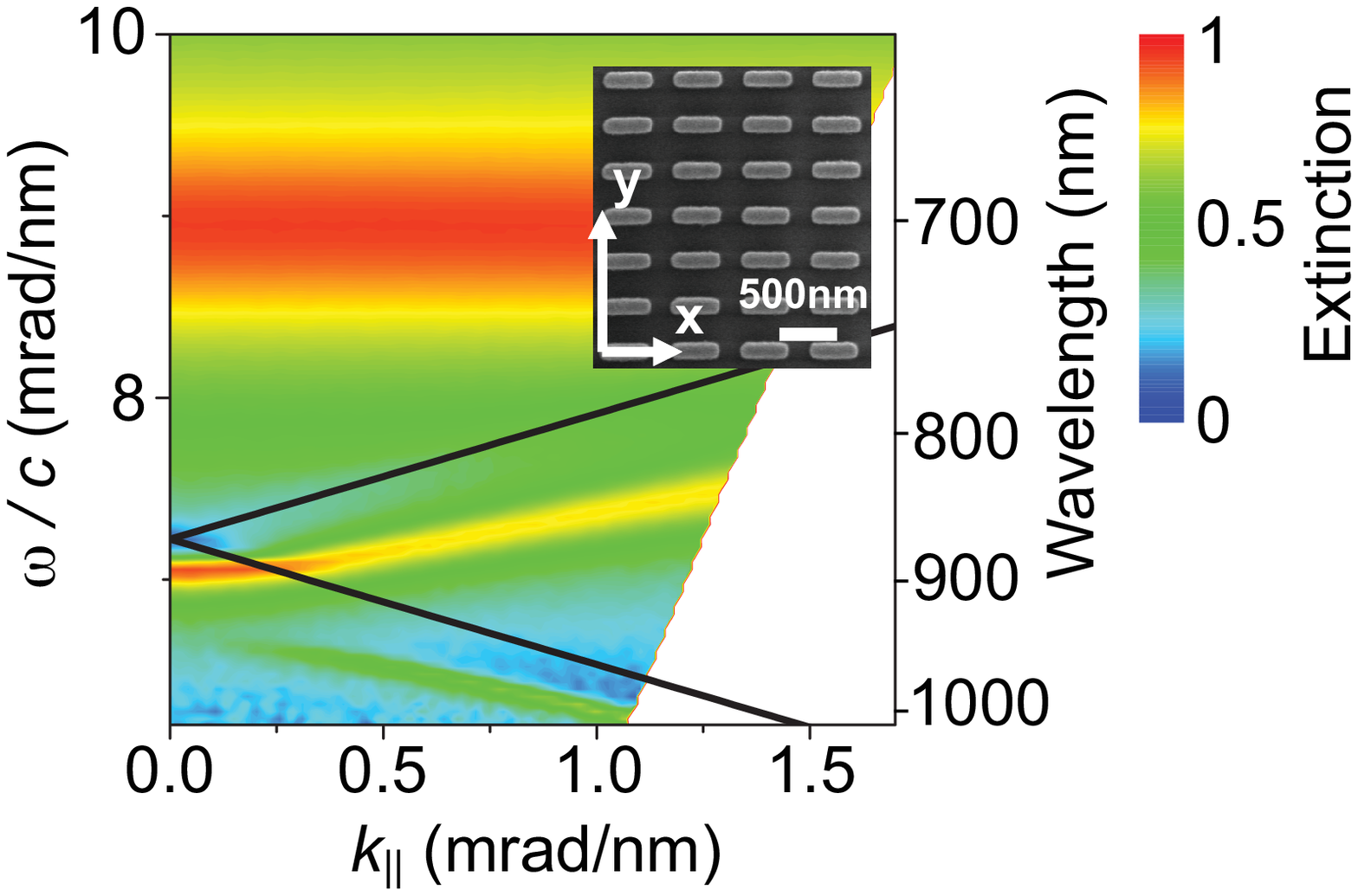}}}
\caption{(Color) Extinction spectra as a function of $k_\|$ for
y-polarized light incident on the array of gold nanorods shown in
the inset. The black lines with positive and negative slope indicate
the $(+1,0)$ and $(-1,0)$ Rayleigh anomalies, respectively. The
dispersionless resonance at 9 mrad/nm is the dipolar localized
surface plasmon resonance for the short-axis of the nanorods,
whereas the narrower and dispersive resonances below the Rayleigh
anomalies are the surface lattice resonances. }\label{fig1}
\end{figure}

 We have investigated  $3 \times 3$ mm$^2$ arrays of gold nanorods fabricated on a silica substrate using substrate conformal
 imprint lithography (SCIL)~\cite{scil}. A top view SEM image of an
 array is displayed in the inset of Figure~\ref{fig1}. This
 array has rods with dimensions $450 \times 120 \times 38$ ${\rm
 nm^3}$ arranged in a lattice with constants $a_x=600$ nm and
 $a_y=300$ nm. The array was embedded in a uniform surrounding
 medium by placing a silica superstrate preceded with n=1.45 index
 matching fluid to ensure good optical contact. We measured the
 variable angle transmittance of the collimated beam from a halogen
 lamp while rotating the sample around the y-axis. The polarization
 of the incident light was set along the y-axis, probing the short axis of the
 nanorods. For this polarization the dipolar LSPR lies at higher energies than the ($\pm 1,0$) diffraction
 orders, thereby allowing the coupling of localized surface plasmons to these orders~\cite{Auguie&Barnes08}.

Figure~\ref{fig1} displays the extinction of the array defined as
$1-T$, with $T$ the transmittance, as a function of the reduced
frequency, i.e., the angular frequency normalized by the speed of
light in vacuum, and the projection of the incident wave vector onto
the surface of the array $k_\|=\frac{\omega}{c}\sin(\theta)
\hat{x}$, with $\theta$ the angle of incidence. The broad,
dispersionless extinction peak centered at a frequency near $9$
mrad/nm corresponds to the excitation of LSPRs in the individual
nanorods. The sharp, dispersive peaks at lower frequencies
correspond to the excitation of SLRs. The ($\pm 1,0$) Rayleigh
anomalies, which are the conditions for which the ($\pm 1,0$)
diffracted orders are grazing to the surface, are indicated with
solid lines in Figure~\ref{fig1}. The coupling of localized surface
plasmons to the Rayleigh anomalies is the origin of the observed
SLRs.

A salient feature in the measurements of Figure~\ref{fig1} is the
formation of a stop-gap centered at $\omega/c = 6.85$ mrad/nm and
near $k_\|=0$, where the two SLRs mutually couple. We note that this
is not a complete photonic bandgap, since it exists for y-polarized
light only. The gap arises from the coupling of two
counter-propagating surface polaritons which, due to the structural
anisotropy of both the nanorods and the lattice, have a strong
polarization dependence on their coupling to free space radiation.
At the high frequency band edge, we observe that the dispersion of
the (+1,0) SLR flattens. This flattening of the band can be
translated as a reduction of the mode's group velocity and the
formation of standing waves, which are also associated with an
increased density of optical states. At the low  frequency band
edge, the (-1,0) SLR becomes weaker and narrower. This behavior is
characteristic of a mode tending towards subradiance, where
radiative damping is suppressed in a collective state with an
antisymmetric wave function~\cite{Dicke54}. As shown in nanoslit
arrays~\cite{Ropers}, there is an intimate connection between
subradiant damping and the opening of a gap in the dispersion
relation of surface plasmon polaritons (SPPs).

\begin{figure}
\centerline{\scalebox{0.5}{\includegraphics{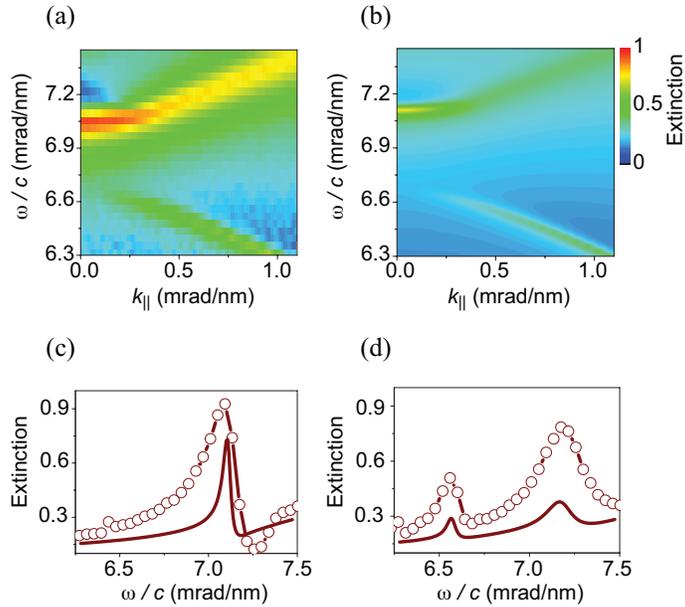}}} \caption{
(Color) Measurements (a) and finite element simulations (b) of the
extinction spectra of the array shown in Figure 1. Figures (c) and
(d) are cuts at $k_\|=0$ and $k_\|=0.4$ mrad/nm, respectively, of
both (a) and (b) . The open circles in (c) and (d) are measurements
and the solid curves are simulations.}\label{fig2}
\end{figure}

Figure~\ref{fig2}(a) shows a close view of the stop-gap in
Figure~\ref{fig1}, and Figure 2(b) shows results from finite element
simulations (COMSOL). For the simulations we used a constant
refractive index of 1.45 for silica, the permittivity of gold as
given in Ref.~\cite{Palik91}, and Bloch-Floquet boundary conditions.
The transmittance was calculated as the ratio between the
transmitted power through the array and the incident power. In
Figures~\ref{fig2}(c) and (d) we compare simulations with
measurements  at $k_\|=0$ and $k_\|=0.4$ mrad/nm, respectively. By
reproducing the measured dispersion of SLRs and the gap's central
frequency and width, a good qualitative agreement between
measurements and simulations is demonstrated. Discrepancies in the
amplitude and spectral width of the resonances can be  mainly
attributed to differences between the simulated and fabricated
geometries, especially near the corners.

It can be appreciated in Figures~\ref{fig1} and~\ref{fig2} that the
SLR peaks together with the Rayleigh anomaly dips give rise to
asymmetric resonance lineshapes, which can be understood in the
framework set forth by Fano~\cite{Fano}. Fano described the quantum
interference between a discrete state and a continuum of states as
the origin of asymmetric resonance lineshapes. It was later shown in
metallic subwavelength hole arrays that the coupling of surface
plasmons to Rayleigh anomalies leads to similar lineshapes, with the
resonantly scattered light acting as the discrete state and the
background transmission as the continuum~\cite{Genet03}. A similar
situation is observed in our configuration. The broad dipolar LSPR
determines the extent of background transmission, i.e. the
continuum, according to its frequency difference with the Rayleigh
anomaly.  The Rayleigh anomaly corresponds to the discrete state.
The diffractive coupling of localized surface plasmons therefore
resembles the interaction between a continuum of states and a
discrete state, leading to asymmetric resonance line shapes. As seen
in Figures~\ref{fig1} and~\ref{fig2}, the degree of asymmetry of
these line shapes changes depending on the frequency and k-vector of
excitation. This dependency is rooted in the relative contributions
of resonant and non resonant scattering~\cite{Galli09}, which also
manifest as a modification of the SLR damping; the latter point will
be addressed further in the text.

\begin{figure}
\centerline{\scalebox{0.5}{\includegraphics{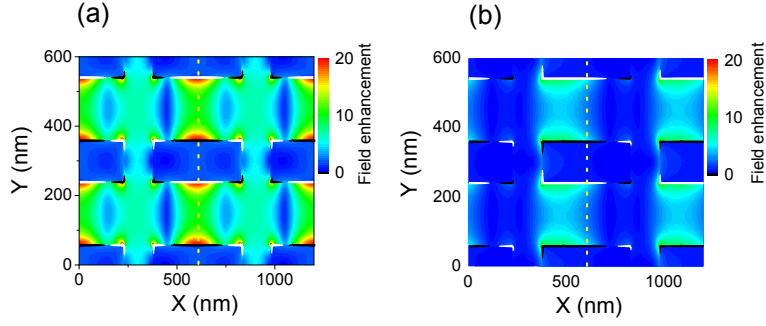}}}
\caption{(Color) Near field enhancement in color scale and surface
charge distribution (at an arbitrary phase) in black and white at
the mid-height of the nanorods for the (+1,0) (a) and (-1,0) (b)
surface lattice resonance at $k_\|= 0.4$ mrad/nm. The dotted lines
indicate the plane of symmetry for coupling to
radiation.}\label{fig3}
\end{figure}

The different electrodynamic response leading to the bright and dark
character - efficient and inefficient coupling to light - of the
(+1,0) and (-1,0) SLRs, respectively, transpires from the near-field
enhancement and surface charge distribution of the nanorods at the
respective frequencies. In Figure~\ref{fig3} we show simulation
results for $k_\| = 0.4 $ mrad/nm at two frequencies: $\omega/c =
7.1 $ mrad/nm and $\omega/c = 6.7$ mrad/nm, which correspond to the
(+1,0) and (-1,0) SLR, respectively. The small angle of incidence
was chosen such that the extinction is not negligible for the (-1,0)
SLR. Both plots are at a plane parallel to the array located at the
mid-height of the nanorods. Charges of opposite sign at the surface
of the nanorods are plotted in white and black, while the total
near-field enhancement, i.e., $|E|^2/|E_0|^2$ with $E$ the total
field and $E_0$ the incident field, is displayed by the color scale.
Figure~\ref{fig3} illustrates how the different resonant response
has its origin in the symmetry of the modes. In order to couple to
the incident plane wave at normal incidence, the mode has to be
symmetric with respect to the plane defined by the incident k- and
polarization vectors intersecting the nanorods along their center,
i.e. the symmetry plane indicated by the dotted lines in
Figure~\ref{fig3}. The (+1,0) mode, which is shown in
Figure~\ref{fig3}(a) for a small angle of incidence, has symmetric
field and charge distributions with respect to the symmetry plane. A
strong dipole moment is seen for each nanorod, and a strong dipolar
inter-rod coupling takes places along the y-direction also. This
results in a large extinction and an efficient coupling of the
(+1,0) mode to normal incidence light, as it can be appreciated in
Figure~\ref{fig1}. In contrast, the (-1,0) mode has an antisymmetric
field and charge distribution for $k_\|=0$; the net dipole moment is
therefore zero and the extinction vanishes. This symmetry is broken
for angles of incidence larger than $\theta= 0^\circ$, thus allowing
the excitation of this resonant mode as shown in
Figure~\ref{fig3}(b). The broken symmetry manifests as a quadrupolar
surface charge distribution displaced from the symmetry axis. This
results in a nonzero intra-rod and inter-rod dipole moment, which
can be recognized from the charges of opposite sign inside the
nanorods and for adjacent nanorods along the symmetry axis,
respectively.

The coupled nature of SLRs can be elucidated by making an analogy
with a set of three mutually coupled harmonic oscillators. Coupled
oscillators have proven useful in understanding electromagnetic
phenomena~\cite{Alzar05, Halas10}. In this analogy, the conduction
electrons in the nanorods driven by the electromagnetic field are
modeled as oscillator 1 driven by a harmonic force $F = F_0
e^{-i\omega_s t}$, whereas the (+1,0) and (-1,0) Rayleigh anomalies
are modeled by oscillators 2 and 3, respectively. The equations of
motion for the system are
 \begin{align}
 \ddot{x_1} + \gamma_1 \dot{x_1} + \omega_1^2 x_1 - \Omega_{12}^2 x_2  - \Omega_{13}^2 x_3 &= F, \nonumber \\
 \ddot{x_2} + \gamma_2 \dot{x_2} + \omega_2^2 x_2 - \Omega_{12}^2 x_1 - \Omega_{23}^2 x_3 &= 0, \\
 \ddot{x_3} + \gamma_3 \dot{x_3} + \omega_3^2 x_3  - \Omega_{13}^2 x_1 - \Omega_{23}^2 x_2 &= 0, \nonumber
 \end{align}

where $x_j$, $\gamma_j$, and $\omega_j$ ($j = 1,2,3$) are the the
displacement from equilibrium position, damping, and eigenfrequency
associated with the $j^{th}$ oscillator, respectively, and
$\Omega_{jk}$ ($k=1,2,3$ and $j\neq k$) is the coupling frequency
between the $j^{th}$ and $k^{th}$ oscillator. Since we are
interested in the extinct optical power in driving the electrons in
the nanorod, we calculate the absorbed mechanical power by
oscillator 1 from the driving force, which is given by $P(t) = F
\dot{x_1}$. Integrating $P(t)$ over one period of oscillation and
scanning the driving frequency $\omega_s$ yields an absorbed power
spectrum $P(\omega_s)$, which is representative of the extinction
spectrum.

In Figure 4 we compare $P(\omega_s)$ with the measured extinction
spectra at three values of $k_\|$. In all three cases $\omega_1 =
9.25$ mrad/nm and $\gamma_1 = 2.3$ mrad/nm, which reproduce the
frequency and damping of the LSPR, and $F_0 = 0.57$ units of force
per mass is a fitting parameter determining the amplitude of the
spectra. The eigenfrequencies $\omega_2$ and $\omega_3$ are
determined for each value of $k_\|$ by the corresponding $(\pm 1,0)$
Rayleigh anomalies of the array. The remaining parameters, i.e.,
coupling and damping frequencies associated with the lattice modes,
are the parameters used to fit the measured SLR lineshapes and are
given in Table 1 for three values of $k_\|$.

\begin{table}
\caption{The $\Omega_{jk}$ terms ($j,k=1,2,3$ and $j\neq k$) are the coupling frequencies between the $j^{th}$ and $k^{th}$ oscillators, and the $\gamma_j$ terms
are the damping frequencies associated with the $j^{th}$ oscillator, for the system described by Equation 1. All quantities are given in units of mrad/nm. In the
entries for which a minimum estimate is given, the value in parenthesis represents the value yielding the spectra in Figure 4.}
\begin{ruledtabular}
\begin{tabular}{l|ccccc}
& $\Omega_{12}$  & $\Omega_{13}$  & $\Omega_{23}$  & $\gamma_{2}$ & $\gamma_{3}$\\
 \hline
$k_\|$=0      &2.8 $\pm$ 0.1     &        -       &        -       &       $<$0.01 (0.001)   &        -\\
$k_\|$=0.17  &     3.1 $\pm$ 0.1      &         $<$0.6 (0.1)     &      1.1 $\pm$ 0.1       &    0.02 $\pm$ 0.01      &       0.020 $\pm$ 0.005\\
$k_\|$=0.68  &     3.3 $\pm$ 0.1      &        2.4 $\pm$ 0.1     &       $<$0.7 (0.3)     &    0.06 $\pm$ 0.02   &       0.03 $\pm$ 0.01\\
\end{tabular}
\end{ruledtabular}
\end{table}

\begin{figure}
\centerline{\scalebox{0.5}{\includegraphics{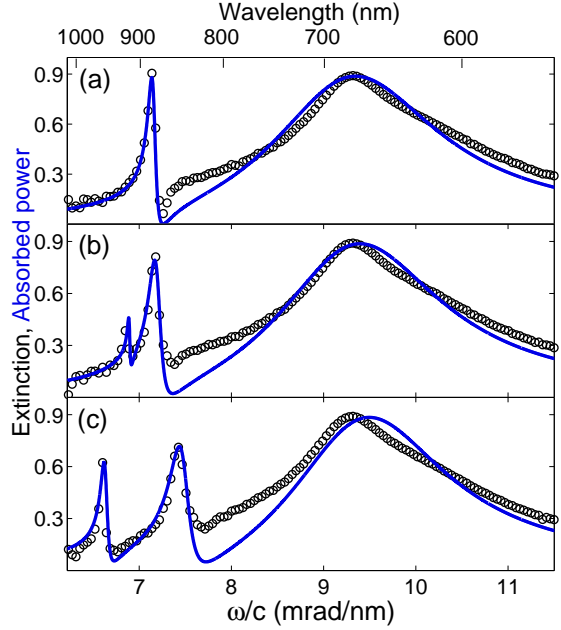}}} \caption{The
black open circles are cuts of the measured extinction spectra shown
in Figure 1 at three values of $k_\|$: (a) $k_\|=0$ mrad/nm, (b)
$k_\|=0.17$ mrad/nm , and (c) $k_\|=0.68$ mrad/nm. The blue solid
curves represent the absorbed power in oscillator 1 of the coupled
oscillator model described in the text, with coupling and damping
frequencies as given in Table 1. }\label{fig4} \end{figure}

Figure~\ref{fig4}(a) displays the spectra at $k_\|=0$,  where the
Rayleigh anomalies are degenerate at $\omega_2 = 7.26$ mrad/nm and
the (-1,0) SLR is a dark state. In this case the model reduces to
that of two coupled oscillators. With increasing $k_\|$ the (-1,0)
SLR comes out of the darkness, so the three oscillators are mutually
coupled. In Figure~\ref{fig4}(b) we consider the case $k_\|= 0.17$
mrad/nm, where $\omega_2 = 7.33$ mrad/nm and $\omega_3 = 6.92$
mrad/nm. From the values given in Table 1 we see that with respect
to the normal incidence case, at $k_\|= 0.17$  mrad/nm the nanorods
are more strongly coupled to the (+1,0) Rayleigh anomaly
($\Omega_{12}$ increases), only weakly coupled to the (-1,0)
Rayleigh anomaly (low $\Omega_{13})$, the Rayleigh anomalies are
mutually coupled (high $\Omega_{23})$, and the damping of both
resonances has very significantly increased (both $\gamma_2$ and
$\gamma_3$ increase). Further increasing to $k_\|= 0.68$ mrad/nm
makes $\omega_2 = 7.68$ mrad/nm and $\omega_3 = 6.70$ mrad/nm, which
is the case in Figure 4(c). At this value of $k_\|$ there is an
increased coupling of the nanorods to both Rayleigh anomalies (both
$\Omega_{12}$ and $\Omega_{13}$ increase), the coupling between the
Rayleigh anomalies decreases ($\Omega_{23}$ decreases), and the
damping of both resonances increases (both $\gamma_2$ and $\gamma_3$
increase).

From the above mentioned behaviors and the quantities given in Table 1 we are able to draw several conclusions. Firstly, the model shows how the coupling terms $\Omega_{jk}$ determine the frequency difference between the SLR peaks and the Rayleigh anomalies as evidenced in Figure~\ref{fig1}. Namely, for the (+1,0) Rayleigh anomaly we observe that as $k_\|$ increases the SLR peak deviates more in frequency. This behavior is defined in the increase of $\Omega_{12}$ which detunes the resonance peak from the Rayleigh anomaly. In contrast, for the (-1,0) diffraction order we observe a decreasing frequency deviation of the SLR from its corresponding Rayleigh anomaly as $k_\|$ increases. In this case the dominant interaction near normal incidence is the mutual coupling of SLRs described by the term $\Omega_{23}$, which detunes the (-1,0) SLR from its Rayleigh anomaly at low $k_\|$. Although $\Omega_{13} > \Omega_{23}$ at large values of $k_\|$, at low values of $k_\|$ the $\Omega_{23}$ interaction dominates due to the smaller frequency difference between the eigenfrequencies $\omega_3$ and $\omega_2$. Secondly, the model shows that the damping of both resonances increases with $k_\|$, which leads to less asymmetric Fano lineshapes and broader linewidths for both resonances. The decreasingly asymmetric lineshapes as $k_\|$ increases are especially clear in Figure~\ref{fig4}, whereas the SLR broadening and variable extinction of the
Rayleigh anomalies are visible in Figure~\ref{fig1}.

It is interesting to calculate the $Q$-factors of the uncoupled
oscillators, which follow from the definition $Q_j = \omega_j /
\gamma_j$. For oscillator 1 resonating at the LSPR frequency we
obtain $Q_{1} = 4$. For oscillator 2 yielding the (+1,0) SLR, we
have $Q_{2} > 700$ at normal incidence. This value, which is to the
best of our knowledge higher than any reported $Q$ for an
experimental study on 2D plasmonic crystals, is a minimum limit.
This limit arises because for any value $\gamma_2<0.01$ a fit to the
measurement within a 10\% uncertainty in the magnitude of the
extinction can be obtained. For oscillator 3 yielding the (-1,0)
SLR, we have $Q_{3} = 300$ at $k_\|=0.17$ mrad/nm. It is important
to realize that the resonances in the coupled system exhibit an
effective damping which is not equal to the damping of the uncoupled
oscillators. Nevertheless, the above values clearly reflect the
large differences in $Q$-factors between LSPRs and SLRs.
Furthermore, the coupled oscillator model points to the origin of
the narrow SLR linewidths, which is the coupling of two harmonic
oscillators with very different damping. By comparing $Q_1$ and
$Q_3$ with previously reported experimental data, we get an insight
into how well the $Q$-factors of the uncoupled oscillators represent
the $Q$-factors of the resonances in the coupled system. Firstly,
$Q_{1} = 4$ is a typical result for LSPRs with high radiative
damping~\cite{Maier}. Secondly, in Ref.~\cite{Ropers} Ropers and
co-workers measured for subradiantly damped SPPs in a nanoslit array
near a stop-gap (similar to what we observe for the (-1,0) SLR near
the gap) lifetimes near $\tau = 200$ fs. Nanoslits can be
represented quite accurately as resonators~\cite{Bozhev06}, so the
lifetime of the excited state, $ \tau$, is related to $Q$ by $Q = 2
\pi \nu \tau$, with $\nu$ the frequency. Inserting $\tau = 200$ fs
and $\nu = 400$ THz (the frequency of the subradiant mode in
Ref.~\cite{Ropers}) yields $Q = 500$, which is comparable with the
value that we find, i.e., $Q_{3} = 300$. Despite the differences in
the structures considered and possibly the extent of subradiant
damping, it is remarkable that the oscillator model yields at the
very least an order of magnitude estimate for the $Q$ of a nearly
subradiant SPP mode in the vicinity of a stop-gap. It should be
mentioned that for arrays of dimensions on the order of 10 microns
or less, the $Q$-factors  of SLRs are expected to appreciably
decrease. This length scale corresponds to the associated surface
polariton propagation lengths, which were recently determined in
Ref.~\cite{Vecchi09b}. For arrays much larger than this
characteristic length, such as the one herein considered, the
$Q$-factor saturates.  The inhibition of radiative damping that
leads to a narrowing of the SLR linewidth is therefore a collective
effect, i.e. it depends on the number of particles present in the
array.

In conclusion, we have demonstrated the coupling of SLRs and the
associated opening of a frequency stop-gap in the dispersion
relation of these modes. The symmetric and antisymmetric
field/charge distributions responsible for the bright and dark
nature of the (+1,0) and (-1,0) modes, respectively, were
illustrated. We have also estimated the very high quality factors of
SLRs ($Q > 700$), which are to the best of our knowledge the highest
reported values for any experimental study on 2D plasmonic crystals.
Coupled bright and dark collective modes, as well as stop-gaps,
offer the possibility to carefully design abrupt changes in the
Local Density of Optical States (LDOS) over narrow spectral regions
and in an extended volume. LDOS manipulation in plasmonic structures
is relevant for enhancing the efficiency of light emitting devices
(LEDs), sensors, and nonlinear processes, all of which can be
tailored in a frequency, angle, and polarization dependent manner
with coupled SLRs. Moreover, the standing waves formed at the  high
frequency band edge hold exciting properties for the manipulation of
light at the nanoscale since they have zero group velocity while
being much less damped than LSPRs. Finally, we envisage that the
strong suppression of radiative losses herein discussed holds great
promise for the development of high $Q$ distributed feedback surface
polariton lasers and plasmonic sensors with enhanced sensitivity.

We thank V. Giannini and M. C. Schaafsma for fruitful discussions,
M. Verschuuren and Y. Zhang for assistance in the fabrication of the
samples. This work was supported by the Netherlands Foundation
Fundamenteel Onderzoek der Materie (FOM) and the Nederlandse
Organisatie voor Wetenschappelijk Onderzoek (NWO), and is part of an
industrial partnership program between Philips and FOM.


%


\newpage

\newpage

\newpage

\newpage 

\end{document}